\documentclass[%
 aip,
 floatfix,
 reprint,
 author-numerical
]{revtex4-1}

\usepackage{graphicx}
\usepackage{amsmath,bm}
\usepackage[labelformat=simple]{subcaption}
\usepackage{multirow}
\usepackage{siunitx}
\usepackage{orcidlink}

\makeatletter
\renewcommand{\p@subsection}{}
\renewcommand{\p@subsubsection}{}
\makeatother
\usepackage{hyperref}
\hypersetup{
     colorlinks=true,
     linkcolor=blue,
     filecolor=blue,
     citecolor=blue,      
     urlcolor=blue,
     breaklinks=true
}

\begin{document}

\title{Hydrodynamics of bubble flow through a porous medium with applications to packed bed reactors}

\author{Pranay P. Nagrani}
\affiliation{School of Mechanical Engineering, Purdue University, West Lafayette, Indiana 47907, USA}

\author{Amy M. Marconnet}
\affiliation{School of Mechanical Engineering, Purdue University, West Lafayette, Indiana 47907, USA}

\author{Ivan C. Christov}
\thanks{Author to whom correspondence should be addressed}
\email{christov@purdue.edu}
\affiliation{School of Mechanical Engineering, Purdue University, West Lafayette, Indiana 47907, USA}

\date{\today}% it is always \today, today

\begin{abstract}
Gas-liquid flows through packed bed reactors (PBRs) are challenging to predict due to the tortuous flow paths that fluid interfaces must traverse. Experiments at the International Space Station showed that bubble and pulse flows are predominately observed under microgravity conditions, while the trickle and spray flows observed under terrestrial conditions are not present in microgravity. To understand the physics behind the former experiments, we simulate bubble flow through a PBR for different packing-particle-diameter-based Weber numbers and under different gravity conditions. We demonstrate different pore-scale mechanisms, such as capillary entrapment, buoyancy entrapment, and inertia-induced bubble displacement. Then, we perform a quantitative analysis by introducing new dynamic scales, dependent upon the evolving gas-liquid interfacial area, to understand the dynamic trade-offs between the inertia, capillary, and buoyancy forces on a bubble passing through a PBR. This analysis leads us to define new dimensionless Weber-like numbers that delineate bubble entrapment from bubble displacement.
\end{abstract}

\maketitle

\section{Introduction}
\label{sec::Intro}

Two-phase flow through a porous medium is a classical topic in fluid mechanics \cite{Adler1988MultiphaseMedia,Drew1983MATHEMATICALFLOW,Wooding1976MultiphaseMedia,Brenner1970RheologySystems,Dias1986NetworkFluids, Koch2001InertialFlows,Avraam1995FlowMedia}. Nevertheless, this topic remains an active research area due to new applications to a variety of systems such as hydrogenation, volatile removal assembly, water treatment plants, and heat pipes, to name a few  \cite{Azarpour2021PerformanceReview,Motil2021GasliquidExperiment,Taghavi2022GasliquidExperiment2}. A packed bed reactor (PBR) is a type of artificial porous medium, comprised of densely packed but randomly distributed catalyst particles within a cylindrical or cuboid column. Commonly, two-phase flows through PBRs involve gas-liquid displacements. The fluids experience a tortuous flow domain as they pass through the PBR, leading to the possibility of a variety of flow regimes in which different physical mechanisms dominate the displacement behavior. Two-phase flow through a PBR under \emph{microgravity} conditions has been of particular interest because, in these conditions, gravity is no longer a dominant force in setting the displacement behavior. Based on experimental characterizations  \cite{Motil2003GasliquidMicrogravity,Motil2021GasliquidExperiment,Taghavi2022GasliquidExperiment2,Taghavi2019GasRate,Deshpande2018EffectHydrodynamics,Zhang2017HydrodynamicsModel} previous work developed empirical models for the streamwise pressure drop, liquid holdup, and flow regime diagram \cite{Salgi2014ImpactBeds,Salgi2015PulseConditions,Salgi2017Experimentally-basedBeds,Taghavi2019GasRate} for gas-liquid flow through PBRs in microgravity conditions. 

% Hydrodynamics %
Motil, Balakotaiah, and Kamotani \cite{Motil2003GasliquidMicrogravity} presented some of the earliest results on two-phase gas-liquid flow through a PBR under microgravity conditions. They observed bubble flow (gas bubbles in the continuous liquid) and pulse flow (alternating flow of gas- and liquid-rich phases) regimes. The typical trickle flow (liquid films clinging to the packing in an otherwise gas-continuous flow) and spray flow (liquid droplets in a continuous turbulent gas phase) regimes observed at terrestrial gravity conditions were not seen in microgravity. Further, by modifying the traditional Ergun equation for single-phase flow \cite{Ranade2011TrickleApplications} through a PBR,  Motil, Balakotaiah, and Kamotani \cite{Motil2003GasliquidMicrogravity} developed a correlation for the frictional pressure drop in gas-liquid two-phase flow. They observed larger pressure drops in microgravity, which they attributed to the dominance of capillary forces in microgravity conditions. Further work in the viscous-capillary regime (\textit{i.e.}, negligible flow inertia) at microgravity was conducted by Motil \textit{et al.}\ \cite{Motil2021GasliquidExperiment} for both wetting and nonwetting packing materials. They found that the capillary contribution dominates the overall pressure drop in the wetting case, while the viscous contribution is dominant for the nonwetting case. Bubble-to-pulse flow regime transition maps were developed for wetting and nonwetting packing materials. In general, Motil \textit{et al.}\ \cite{Motil2021GasliquidExperiment} observed that, as the gas flow rate is increased for a given liquid flow rate, the regime transitions from bubble flow to pulse flow. Taghavi \textit{et al.}\ \cite{Taghavi2022GasliquidExperiment2} extended the previous studies of gas-liquid flows in PBRs under microgravity conditions to longer-duration runs and smaller packing diameters, and they considered different gas and liquid pre-flow conditions. In addition to the dispersed bubble and pulse flow regimes, Taghavi \textit{et al.}\ \cite{Taghavi2022GasliquidExperiment2} also observed regimes of large bubble flow and gas channeling. Due to the smaller packing diameter, the influence of capillary forces and, hence the overall pressure drop, was larger than that measured in the earlier experiments by Motil \textit{et al.}\ \cite{Motil2003GasliquidMicrogravity,Motil2021GasliquidExperiment}. More recently, Zhang \textit{et al.}\ \cite{Zhang2017HydrodynamicsModel} used a combination of experiments and two-fluid modeling to understand the pressure drop and liquid holdup trends with varying gas and liquid flow rates for different bead (or packing) sizes and liquid properties. An increase in the pressure drop and liquid holdup was observed for smaller bead sizes.

Overall, these prior works on the hydrodynamics of gas-liquid flow through a PBR under microgravity (or microgravity-like) conditions broadly focused on flow regime characterization (into a bubble or pulse flow) and the corresponding pressure drops, without gaining insight into the bubble dynamics. Moreover, in the recent analysis of the microgravity experiments on the ISS, Motil \textit{et al.}\ \cite{Motil2021GasliquidExperiment} noted that ``one fundamental problem identified in the PBRE, that is, there is a minimum liquid superficial flux needed to dislodge trapped bubbles in a porous medium.'' To better understand this fundamental problem and thus fill a knowledge gap in the literature, we explore the bubble flow regime in depth by investigating the influence of inertia, capillary, and buoyancy forces on a bubble's entrapment and displacement through a PBR. To this end, we propose new dynamic scales to estimate the pore-scale force magnitudes and balances on a bubble traversing a PBR, which improves on the assumption of previous studies  \cite{Motil2003GasliquidMicrogravity,Motil2021GasliquidExperiment,Taghavi2022GasliquidExperiment2} that the dimensionless numbers should be defined based on a constant packing diameter, without taking into consideration the scales characterizing the tortuous paths through the PBR. 

% PBR CFD %
Our study of bubble dynamics is based on computational fluid dynamics (CFD) simulations, which offer insights into the complex flow physics of PBRs, beyond what can be measured or observed in experiments. 
In particular, interface evolution algorithms, such as the volume-of-fluid (VOF) method, have widely been used to understand the flow regimes and their transition in PBRs \cite{Ranade2011TrickleApplications}. In this vein, previous CFD studies simulated gas-liquid flow through PBRs under terrestrial gravity conditions. For example, pore-scale-resolved transient 3D simulations of gas-liquid flow through porous media (including PBRs) were performed by Ambekar \textit{et al.}\ \cite{Ambekar2021Pore-resolvedMap,Ambekar2022ForcesBeds,Ambekar2022Particle-resolvedSize} to understand the dynamics of liquid spreading and oil recovery. In particular, Sun and Santamarina \cite{Sun2019HainesMechanisms} and O'Brien, Afkhami, and Bussmann \cite{OBrien2020Pore-scaleModel} used the VOF method to understand pore-scale phenomena such as Haines jumps. Meanwhile, Ambekar, Mondal, and Buwa \cite{Ambekar2021Pore-resolvedMap} investigated the effects of wettability (via the liquid-solid contact angle) and the relative importance of capillary forces on oil displacement patterns and how these quantities set the different pore-scale flow regimes (such as ``finger-like invasion," ``co-operative filling," and ``pore-by-pore filling"). Further, Ambekar, R\"{u}de, and Buwa \cite{Ambekar2022ForcesBeds} proposed a dimensionless number that accounts for the relative magnitude of a combined inertia--capillary force over gravity. It was observed that liquid spreading was dominated by inertial forces at short times and by capillary forces at long times, which results in enhanced lateral spreading. On the other hand, gravitational forces restricted lateral liquid spreading. In addition, a flow regime map was introduced, highlighting bubble and trickle flow regions. This work was further extended by Ambekar, R\"{u}de, and Buwa \cite{Ambekar2022Particle-resolvedSize}, who varied the wettability and packing diameter to study their influence on liquid spreading. They showed that, for small contact angles and packing diameters, the inertial and capillary forces dominate the liquid spreading in the lateral direction, while for larger packing diameters and contact angles, gravitational forces dominate, resulting in the mitigation of lateral liquid spreading. Xu \textit{et al.}\ \cite{Xu2022Particle-resolvedFlow} obtained further insights into the bubble flow regime. Specifically, using a combination of experiments and CFD simulations, they investigated bubble coalescence in a concurrent upward gas-liquid flow (opposing gravity) through a PBR. As the gas superficial velocity was increased, larger-diameter bubbles were observed to coalesce at the bottom of the PBR, resulting in reduced bubble coalescence times followed by increased pressure drops. Similar observations were reported for increasing liquid superficial velocities. 

As noted above, previous CFD studies focused on gas-liquid flow through PBRs under terrestrial gravity conditions and specifically on the flow regimes arising when the liquid phase displaces the gas phase. Going beyond these previous works, we perform 3D transient CFD simulations using the volume-of-fluid (VOF) method to study the bubble dynamics in PBRs under microgravity conditions. To this end, in Section~\ref{sec::model_methods}, we introduce a computational model for gas-liquid flow in a PBR at microgravity conditions. Specifically, we discuss the PBR geometry generation followed by the meshing approach and CFD setup. Then, in Section~\ref{sec::RAndD}, we explore the effect of varying liquid inertia and gravity conditions on the bubble displacement profiles. We propose to understand pore-scale displacements and dominant mechanisms via dynamic balances between the inertia, capillary, and buoyancy forces. Specifically, we introduce new dynamic scales, dependent on the interfacial gas-liquid area, to quantitatively explain the evolution of the force balances. Using volume fraction maps, we show qualitatively the transition from the bubble to the pulse flow regime when two bubbles coalesce in the radial direction to form a pulse. Finally, Section~\ref{sec::Conclusion} concludes the study.

\section{Modeling and simulation methodology}
\label{sec::model_methods}

\subsection{Generation and meshing of the packed bed geometry}
\label{sec:PBR_Geom_Mesh}

The 3D geometry of a PBR consists of a random spherical packing of particles within a cylindrical column. We constructed the 3D geometry by conducting rigid body particle simulation in the 3D modeling and rendering software Blender \cite{BlenderOnlineCommunity2018BlenderPackage}. We placed $1979$ particles of $d_p=3~\si{mm}$ diameter at the top of a cylindrical column of $2R = 30~\si{mm}$ diameter and $75~\si{mm}$ height. The particles were allowed to fall into the cylindrical column due to the force of gravity. When the particle-particle and particle-cylinder contact forces balance gravity, the static packing can be determined. We tuned the simulation parameters (listed in Table~\ref{tab::blender_param}) for the Blender simulation based on prior studies in the literature \cite{Hernandez-Aguirre2022FramingReactor,Boccardo2014Pore-scaleCFD,DP_IntegratedWorkflow}.

\begin{table}
    \setlength{\tabcolsep}{7pt}
    \begin{tabular}{lll}
      \hline
      \hline
      \noalign{\vskip 0.5ex}
      Rigid body simulation parameter  & Value \\
      \hline
      \noalign{\vskip 0.5ex}
      Friction coefficient & $0.9$ \\
      Coefficient of restitution  & $0.9$  \\
      Damping coefficient & $0.1$ \\
      Margin  & $0~\si{\centi\meter}$ \\
      \hline
      \hline
    \end{tabular}
    \caption{Parameters used in Blender to generate the representative 3D PBR geometry as a random spherical packing within a cylindrical column.}
    \label{tab::blender_param}
\end{table} 

\begin{figure}
    \centering\includegraphics[width=\linewidth]{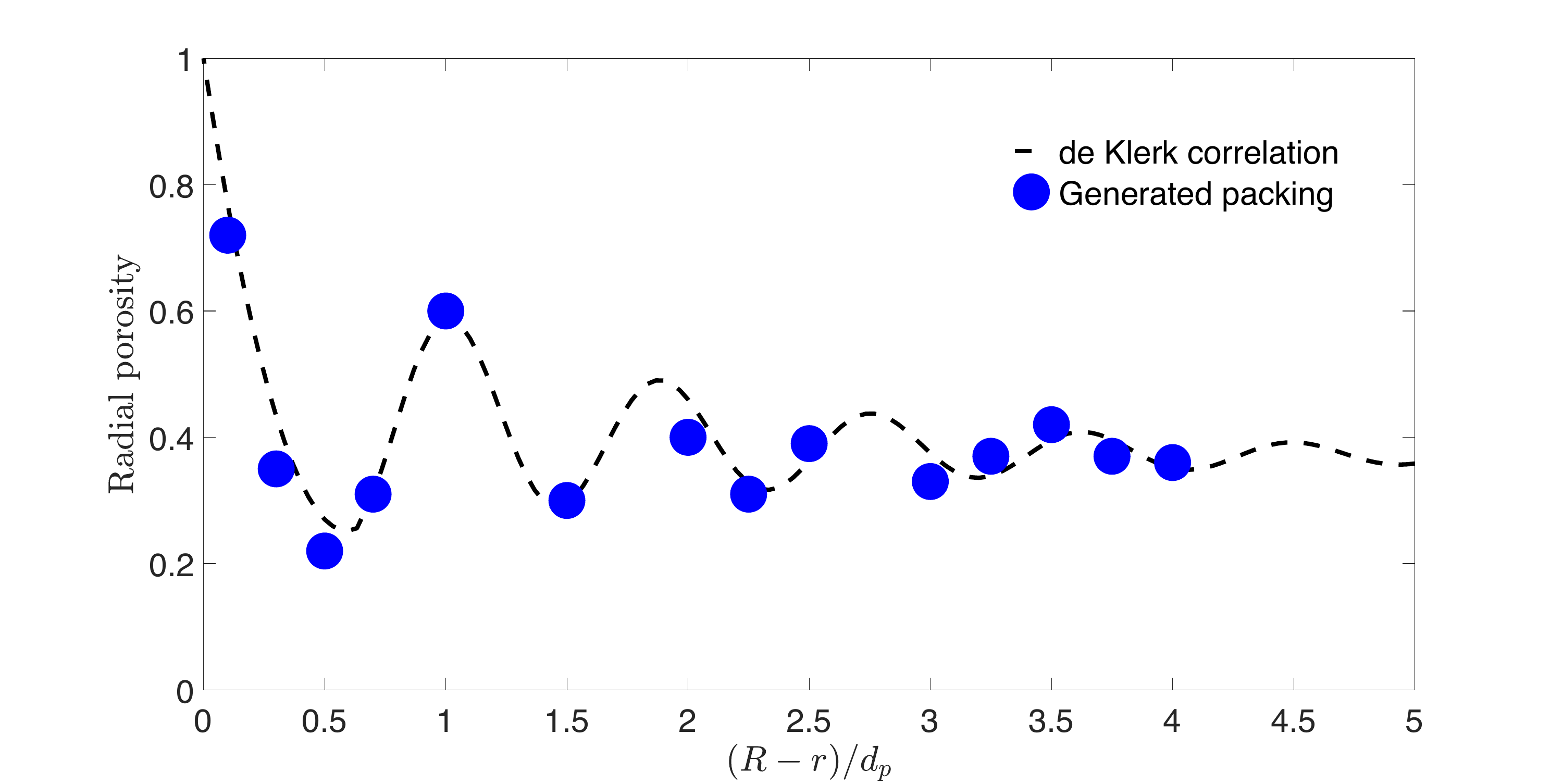}
    \caption{Comparison of the radial porosity variation of the 3D PBR geometry generated for this study and the correlation of de Klerk \cite{DeKlerk_Val}.}
    \label{fig::PBR_Geom_Val}
\end{figure}

\begin{figure}
    \centering\includegraphics[width=\linewidth]{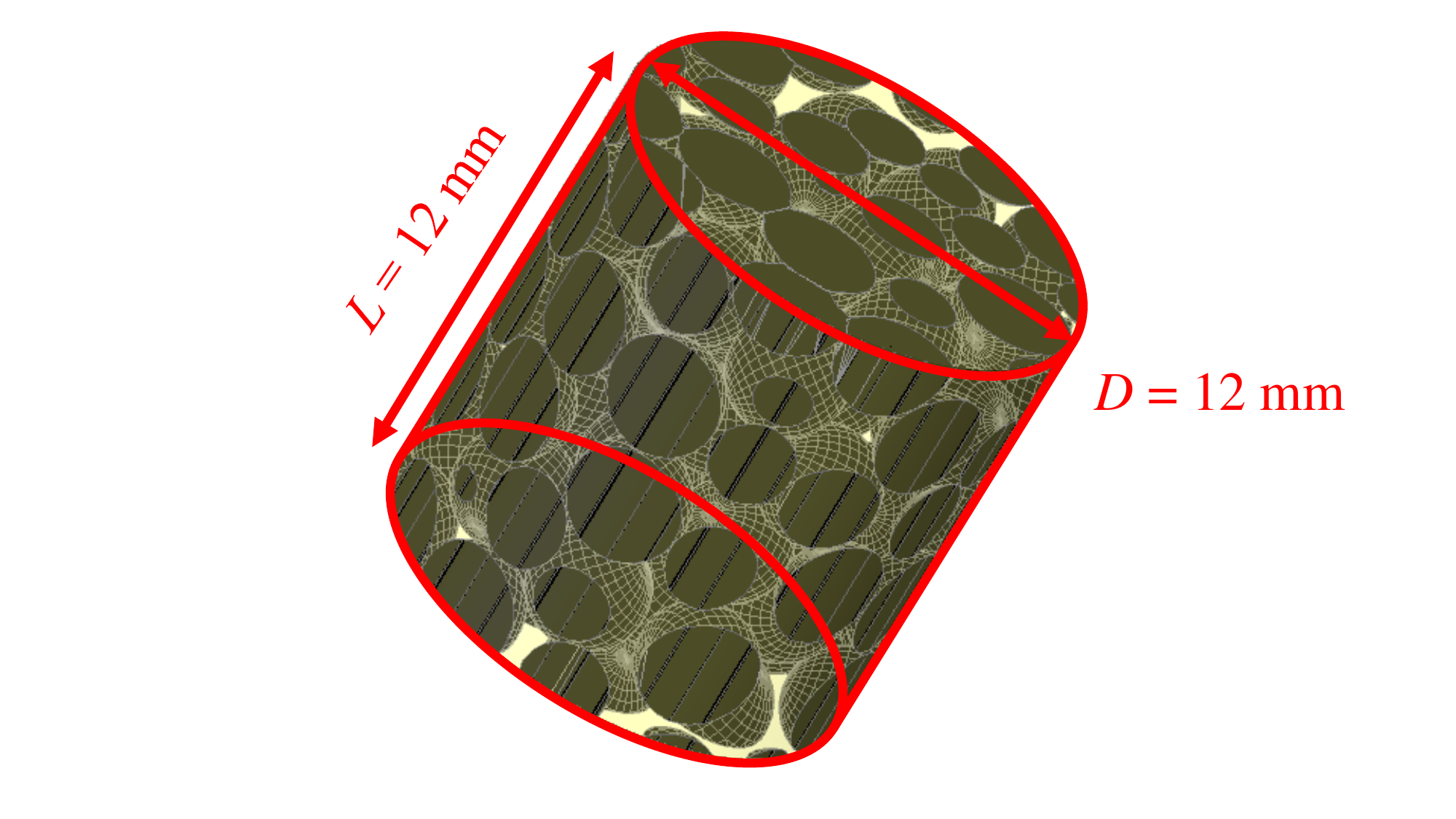}
    \caption{PBR representative element volume (REV) generated for the CFD simulations, comprising of a $3~\si{mm}$ diameter sphere packing within a cylindrical column of $12~\si{mm}$ height and $12~\si{mm}$  diameter. Note that the darker regions in the figure represent the spherical packing.}
    \label{fig::PBR_Geom_PBR}
\end{figure}

We performed further processing on the simulated PBR geometry using ANSYS SpaceClaim 2022 R1 \cite{ANSYS_SC}. Specifically, we discarded the spheres intersecting the bottom and top faces of the cylindrical column to prevent end effects, which can also affect the radial porosity (meaning, the ratio of the area of the pores to the total surface area along the radial direction). Next, we calculated the radial porosity of the modified packing structure and compared it to the analytical correlation of de Klerk \cite{DeKlerk_Val} in Fig.~\ref{fig::PBR_Geom_Val}. This comparison establishes that the packing structure represents a realistic PBR. Finally, we extracted a representative element volume (REV) \cite{Sun2022ConvectiveITheory,Sun2022ConvectiveValidation,Boccardo_2015,Hernandez-Aguirre2022FramingReactor} of the PBR of diameter $12~\si{mm}$ and length $12~\si{mm}$, as shown in Fig.~\ref{fig::PBR_Geom_PBR}. The REV was extracted from the middle of the Blender-generated PBR geometry to ensure the porosity fluctuations are small.

\begin{figure}
    \centering\includegraphics[width=\linewidth]{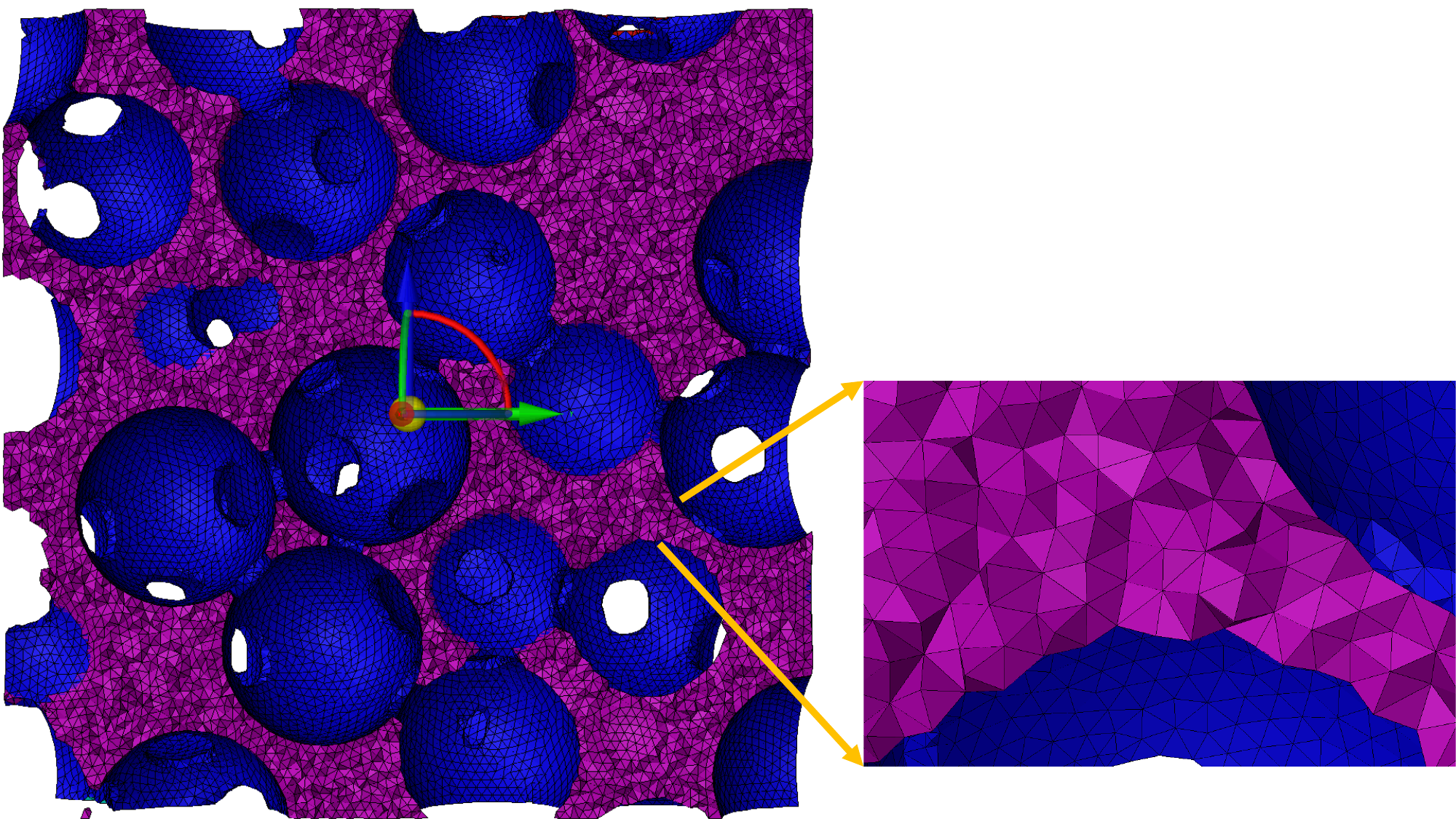}
    \caption{Surface and volume mesh of the PBR REV generated using the fault-tolerant meshing workflow with the parameters in Table~\ref{tab::mesh_param}.}
    \label{fig::PBR_Geom_Mesh}
\end{figure}

In the PBR, the fluid domain is the 3D volume obtained by Boolean subtraction of the sphere packing from the cylindrical column. The resulting fluid domain is meshed into tetrahedral elements for the simulations. Additional treatment of the contact point between two spheres or between a sphere and the cylinder wall is necessary to avoid overly skewed mesh elements. To this end, we used the shrink-wrap algorithm \cite{Lee_shrinkwrapping} implemented in the fault-tolerant meshing workflow in ANSYS Fluent Meshing 2022 R1 \cite{ANSYS_Fluent1}. The parameters used in this fault-tolerant meshing workflow are given in Table~\ref{tab::mesh_param}. The shrink-wrapping meshing approach is typically used for nonwatertight or ``dirty'' computer-aided design (CAD) models (containing holes, gaps, etc). In this approach, an initial surface mesh is generated by wrapping (or enveloping) the spherical particles, resulting in the creation of bridges if the two neighboring spheres are closer than a specified tolerance \cite{DP_IntegratedWorkflow}. These bridges regularize point contacts. To avoid highly skewed cells, we created a surface mesh with a maximum skewness of $0.7$. Next, using the surface mesh as the starting point, we created a volume mesh in the fluid flow domain. We checked the accuracy of the generated mesh by ensuring the volume mesh has a skewness below $0.85$ and orthogonal quality above $0.15$. Further manual checks were performed for ``bad'' cell elements, such as ones with negative volumes, large aspect ratios, etc. The final mesh has approximately $1.3$ million elements, and it is shown in Fig.~\ref{fig::PBR_Geom_Mesh}. In this figure, we highlight the volume mesh of the fluid domain and the surface mesh on the spherical packing. A zoom-in shows the mesh between two spheres, suggesting that sufficient resolution is available to accurately resolve the flow and gas-liquid interface through such tortuous paths. 

\subsection{CFD approach and its implementation}
\label{sec:Gov_Eqns}

We used the VOF method \cite{Hirt1981VolumeBoundaries,FP02} to simulate the gas-liquid dynamics. This method does not separately resolve the gas and liquid phases' dynamics. Rather, the conservation of mass (continuity) and momentum equations are written for a gas-liquid mixture as:
\begin{subequations}\begin{align}
\bm{\nabla\cdot}\bm{v} &= 0, \\
\frac{\partial(\rho \bm{v})}{\partial t} + \bm{\nabla\cdot} (\rho \bm{v} \otimes \bm{v}) &= 
-\bm{\nabla} p + \bm{\nabla\cdot} [\eta (\bm{\nabla}\bm{v} + \bm{\nabla}\bm{v}^\top )] \nonumber  \\
&\qquad + \rho\bm{g} + \bm{f}_\mathrm{csf}, \label{eq:momentum}
\end{align}\label{eq:iNS}\end{subequations}
where $\rho$, $\bm{v}$, and  $\eta$ are the mixture's density, velocity, and dynamic viscosity, respectively, $p$ is the pressure, $\otimes$ denotes the (direct) dyadic product, $\bm{g}$ is the gravitational acceleration vector, and $\bm{f}_\mathrm{csf}$ is a fictitious body force used to enforce surface tension at the gas-liquid interface. In this formulation, all of the latter quantities are functions of $x$, $y$, $z$, and $t$. 

\begin{table}
    \setlength{\tabcolsep}{7pt}
    \begin{tabular}{lll}
      \hline
      \hline
      \noalign{\vskip 0.5ex}
      Mesh parameter  & Value \\
      \hline
      \noalign{\vskip 0.5ex}
      Shrink factor & $0.4$ \\
      Size functions  & Curvature and proximity  \\
      Min.\ size  & $d_p/20$ \\
      Max.\ size & $d_p/10$ \\
      Max.\ skewness & $0.85$ \\
      Min.\ orthogonal quality & $0.15$ \\
      \hline
      \hline
    \end{tabular}
    \caption{Mesh parameters used in ANSYS Fluent Meshing to generate a high-quality mesh for the CFD simulations of flow in a PBR.}
    \label{tab::mesh_param}
\end{table} 

The (dimensionless) phase fractions $\alpha_l$ and $\alpha_g$, where the subscripts ``$g$'' and ``$l$'' henceforth refer to the gas and liquid, respectively, keep track of how much of each fluid is present in each computational cell. For example, $\alpha_l(x,y,z,t) \in (0,1)$ corresponds to locations along the ``diffused'' approximation of the gas-liquid interface, while $\alpha_l(x,y,z,t) = 1$ or $0$ means the point $(x,y,z)$ is within a mesh cell that contains only liquid, and vice versa for $\alpha_g$. The evolution of $\alpha_l$ is determined by the transport equation
\begin{equation}
    \frac{\partial(\rho_l\alpha_l)}{\partial t} + \bm{\nabla\cdot} (\rho_l\alpha_l \bm{v}) = 0,
    \label{eqn::vol_frac_conv}
\end{equation}
assuming no sources and no mass transfer between the fluids. Then, the volume fraction of the gas phase is $\alpha_g = 1-\alpha_l$. The thermophysical properties of the mixture to be used in Eqs.~\eqref{eq:iNS}, are calculated as 
\begin{subequations}\label{eq:mixture}
    \begin{align}
        \rho &= \alpha_g \rho_g + \alpha_l \rho_l,\label{eq:rho_mixture}\\
        \eta &= \alpha_g \eta_g + \alpha_l \eta_l.
        %\bm{v} &= \alpha_g \bm{v}_g + \alpha_l \bm{v}_l.
    \end{align}%
\end{subequations}
Therefore, to fully specify this computational model and study the dynamics of gas-liquid flows, it suffices to know the gas and liquid's thermophysical properties. 

The fictitious force, $\bm{f}_\mathrm{csf}$, in Eq.~\eqref{eq:momentum} arises because surface tension causes a pressure jump at gas-liquid interfaces (Young--Laplace law). This interfacial force is distributed as a body force (per unit volume) in the mixture model~\eqref{eq:iNS} and evaluated using the continuum surface force (CSF) method \cite{Brackbill1992ATension,Popinet2018NumericalTension,ANSYS_Fluent2} as 
\begin{equation}
    \bm{f}_\mathrm{csf} = \sigma \frac{\rho}{(\rho_l+\rho_g)/2} \kappa \bm{\nabla} \alpha_l,
\end{equation}
where $\sigma$ is the gas-liquid surface tension, $\kappa = \bm{\nabla\cdot}\bm{n}$ is the mean curvature of the gas-liquid interface computed directly from its shape via the surface normals $\bm{n}=\bm{\nabla}\alpha_l/\|\bm{\nabla}\alpha_l\|$, and $\rho$ is computed as in Eq.~\eqref{eq:rho_mixture}. Note that from $\alpha_l+\alpha_g=1$, it follows that $\bm{\nabla}\alpha_l=-\bm{\nabla}\alpha_g$, and the VOF method can be equivalently formulated using $\alpha_g$ instead of $\alpha_l$.

\subsection{Simulation methodology}
\label{sec:comp_method}
% Model setup details in Fluent

To understand bubble dynamics under microgravity conditions, we performed transient 3D CFD simulations of the bubble regime of gas-liquid flow in a PBR using the VOF method implemented in ANSYS Fluent 2022 R1 \cite{ANSYS_Fluent2}. We take air as the gas phase and water as the liquid phase. The material properties are listed in Table~\ref{tab::props}. In experiments on two-phase flow through a PBR, the gas and liquid phases are injected at fixed flow rates at the inlet, which in turn determines the prevalent flow regime that will be observed in the PBR \cite{Motil2003GasliquidMicrogravity,Motil2021GasliquidExperiment,Taghavi2022GasliquidExperiment2}. Since a simulation of an entire PBR is computationally expensive, we simulated bubble flow in the PBR REV that we constructed in Section~\ref{sec:PBR_Geom_Mesh}. To access the bubble flow regime, at $t=0~\si{\second}$ we patched a $5~\si{mm}$ diameter air bubble at the top of the REV and initialized the rest of the flow domain with water at rest ($\alpha_l=1$). 

\begin{table}
    \setlength{\tabcolsep}{7pt}
    \begin{tabular}{llll}
      \hline
      \hline
      \noalign{\vskip 0.5ex}
      Phase & $\rho~(\si{\kilo\gram\per\meter\cubed})$ & $\eta~(\si{\pascal\second})$ & $\sigma~(\si{\newton\per\meter})$\\
      \hline
      \noalign{\vskip 0.5ex}
      Air & $1.225$ & $1.79\times10^{-5}$ 
      & \multirow{2}{*}{0.072} \\
      Water & $998.0$ & $1.03\times10^{-3}$ &  \\
      \hline
      \hline
    \end{tabular}
    \caption{Material properties of air (gas phase) and water (liquid phase) used in the CFD simulations.}
    \label{tab::props}
\end{table} 

\begin{figure}
    \centering
    \includegraphics[width=\linewidth]{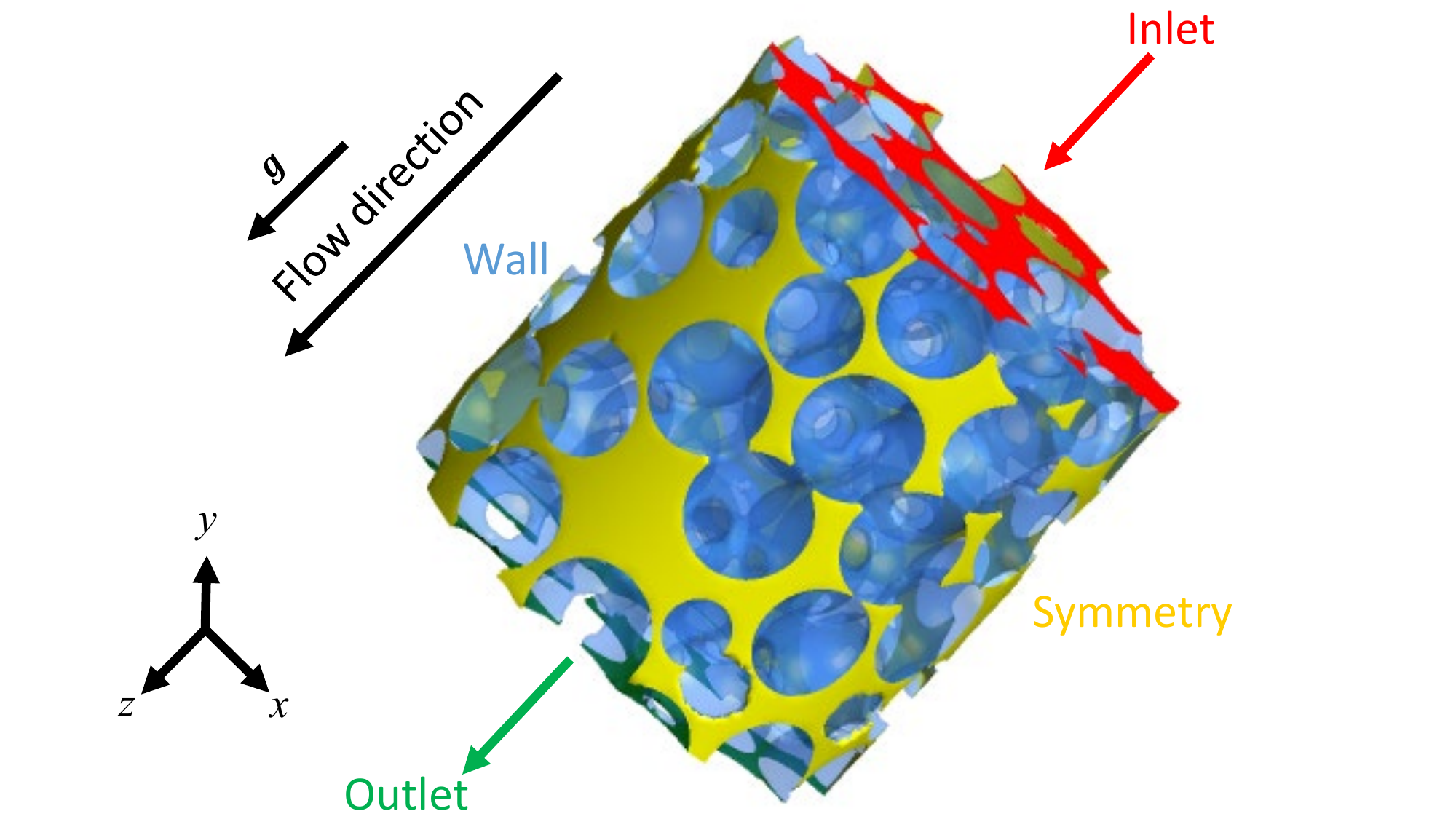}
    \caption{Schematic representation of the REV geometry and boundary conditions on it used in the CFD simulations of the dynamics of bubble flow through a PBR.}
    \label{fig::BC}
\end{figure}

The boundary conditions for our study are summarized in Fig.~\ref{fig::BC}. On the inlet plane, we specified a velocity boundary condition as well as $\alpha_l=1$. On the outlet plane, we specified $0~\si{\pascal}$ gage pressure. On the spherical packing walls, we specified a no-slip velocity boundary condition. Meanwhile, the outer cylindrical column wall was defined as a symmetry plane to implement the REV concept used in the current analysis, and since we did not simulate an entire PBR. We did not consider the effect of wettability in the present study, so contact angle dynamics were not modeled. We expect that the influence of the contact angle on flow dynamics is negligible in liquid continuous flow regimes, such as the bubble flow studied in this work \cite{Motil2021GasliquidExperiment}. In Section~\ref{sec:grav}, unless otherwise stated, microgravity conditions ($g=10^{-4}~\si{\meter\per\second^{2}}$) are considered in the simulations.

In ANSYS, we used a ``Least Squares Cell Based" scheme to discretize gradients. The pressure-velocity coupling in the momentum equation was achieved using the pressure implicit splitting of operators (PISO) algorithm. In this context, the pressure-correction and momentum equation \eqref{eq:momentum} were discretized using the ``Body Force Weighted" and ``Second Order Upwind" numerical schemes. The ``Geo-Reconstruct" algorithm was used to reconstruct a sharp gas-liquid interface, with reduced numerical diffusion, and solve the transport equation \eqref{eqn::vol_frac_conv}. For more information, the reader is referred to the ANSYS Fluent User's Guide \cite{ANSYS_Fluent1}. The simulations were run with a time step of $\Delta t = 3 \times 10^{-6}~\si{\second}$ to maintain a global Courant number less than $0.5$ in the explicit VOF-based solver for the transport equation. The continuity and momentum residuals were ensured to achieve relative values of $1\times10^{-4}$ in each time step. Each simulation was run on $24$ cores on a parallel computing cluster, which took roughly $2$ weeks to complete. To be consistent across simulations, we introduce the scaled (dimensionless) time $t^*=t/t_\mathrm{end}$, where $t_\mathrm{end}=0.15~\si{\second}$ is the end time of a simulation. 

We performed a grid-convergence study to verify our simulations and to determine the optimum mesh size required to perform high-fidelity CFD simulations while keeping the computational time reasonable. We used three different meshes in the grid-convergence study -- a coarse mesh with $94\,582$ elements, a medium mesh with $1\,281\,148$ elements (used in the present study), and a fine mesh with $2\,451\,978$ elements. These meshes were obtained by varying the minimum and maximum element sizes in ANSYS Fluent Meshing 2022 R1 \cite{ANSYS_Fluent1}. For the coarse mesh, the bridges formed as the bubble navigates the pore spaces are large, and the gas-liquid interface is diffuse in nature. On the other hand, for the medium and fine meshes, the bridges formed are of the same size, and the interface is sharp, indicating that the medium mesh used in the present study can accurately capture various pore-scale phenomena.

\section{Results and discussion}
\label{sec::RAndD}
In this section, we introduce novel, time-dependent scales to capture the pore-scale effects that set the evolution of the forces acting on a bubble in a PBR. Next, we discuss the bubble dynamics for different values of the global Weber number $We_{d_p} = \rho_l U_{ls}^2 d_p/\sigma$ based on the spherical packing's diameter $d_p$. We obtain three values of $We_{d_p}$ by varying the liquid phase's superficial velocity $U_{ls}$. The superficial velocity is the mean velocity across a cross-sectional plane, calculated here by dividing the phase's inlet flow rate by the inlet area. Further, for one value of $We_{d_p}$, we vary the magnitude of the gravitational acceleration. We use the pore-resolved CFD simulation approach introduced in Section~\ref{sec::model_methods} to track the air-water interface in a PBR REV under these conditions.

\begin{figure*}
    \centering
    \begin{subfigure}[t]{\linewidth}
      \includegraphics[width=\linewidth]{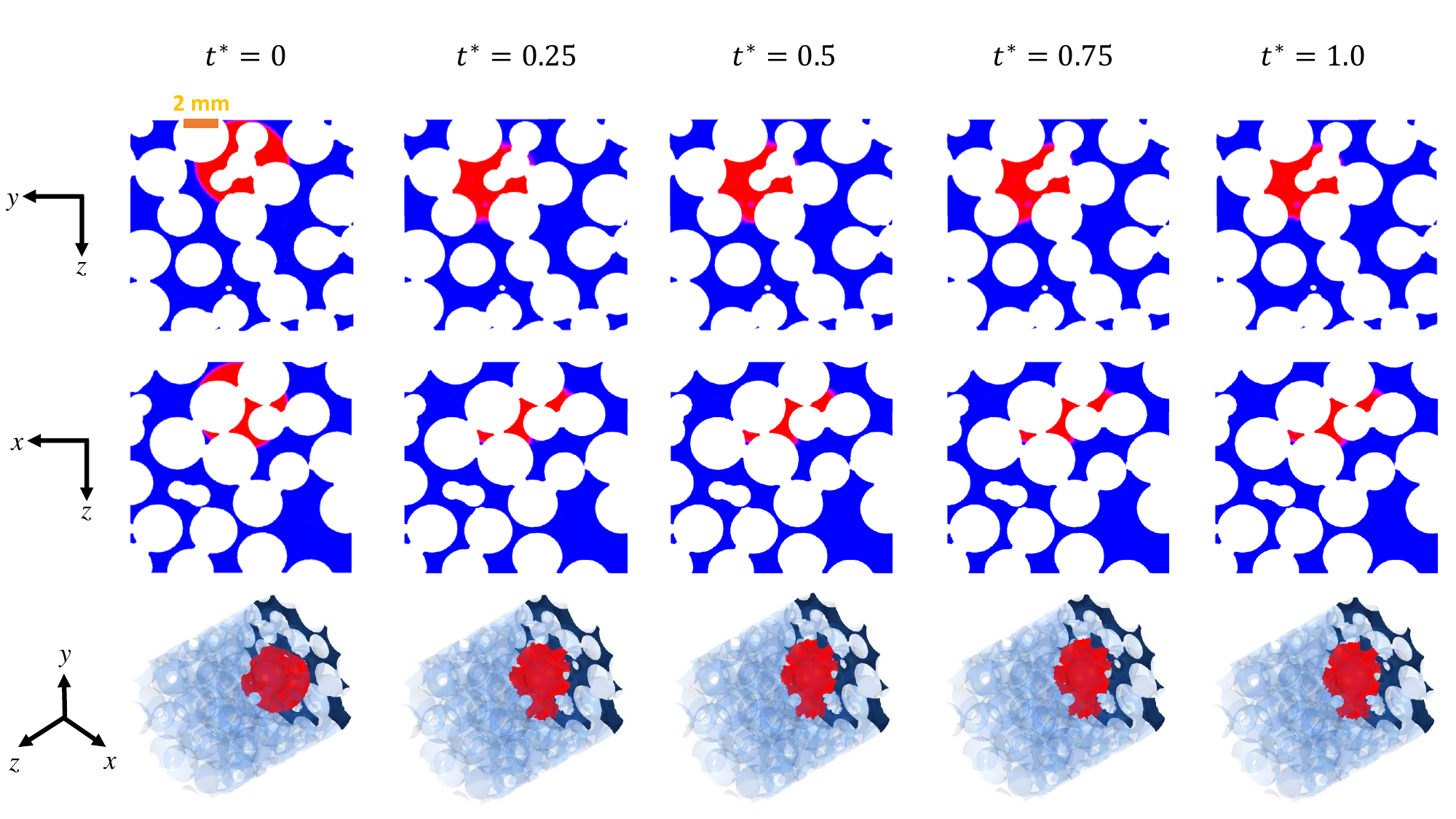}  
      \caption{}
      \label{fig::Uls_0.063ms_0g}
    \end{subfigure}
    \begin{subfigure}[t]{\linewidth}
      \includegraphics[width=\linewidth]{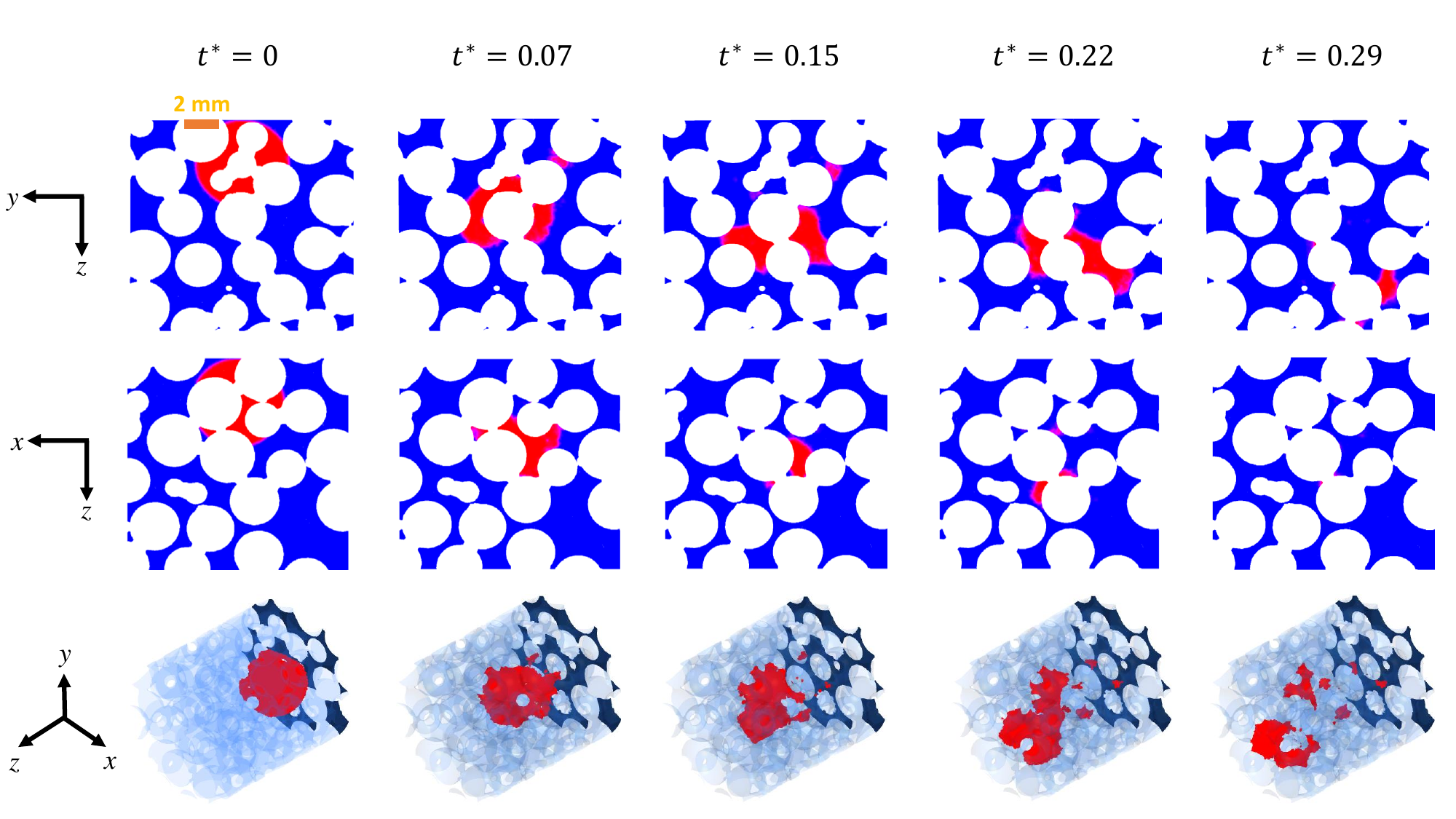}  
      \caption{}
      \label{fig::Uls_0.155ms_0g}
    \end{subfigure}
    \caption{Bubble evolution Bubble evolution in 2D axial profiles (top and middle rows) and 3D isometric views (bottom rows) for (a) $We_{d_p} = 0.165$ showing capillary entrapment and (b) $We_{d_p} = 1.0$ showing inertia-induced bubble displacement. Note the liquid flow is in the $+z$ direction.}
    \label{fig::Uls_0.1ms_0.155ms_0g}
\end{figure*}

\subsection{Introduction of suitable scales}
We propose the interfacial area $A_\mathrm{int}$ as the suitable dynamic scale to use in determining the magnitude, $F_{I}$, of the inertia force. From $A_\mathrm{int}$, we infer a self-consistent dynamic length scale, $A_\mathrm{int}^{1/2}$, to estimate the magnitude of the buoyancy force $F_{B}$. Specifically, in ANSYS Fluent we define an instantaneous surface $\{ (x,y,z) \,|\, \alpha_l(x,y,z,t)=0.5\}$, which corresponds to the gas-liquid interface. Then, using a built-in function, we compute the area of this surface, yielding $A_\mathrm{int}(t)$. This choice of dynamic scale allows for the evolution of $F_{I}$ and $F_{B}$ with the flow by taking into account the pore-scale characteristics. We show below that capturing the dynamic nature of the forces (and force balances) is critical to understanding the hydrodynamics of the bubble flow. Further, $F_{I}$ is defined based on the area-averaged interfacial velocity $U_\mathrm{int}$. We estimate the magnitude, $F_{C}$, of the capillary force acting on the bubble by using the circumference of the bubble, $\pi d_B$, as the length scale. This estimate is based on the static bubble configuration, consistent with the literature, because the Young--Laplace law from which it is inferred concerns only fluid statics. To summarize, we estimate the magnitudes of the three forces acting on the bubble in a PBR as: 
\begin{subequations}
    \begin{align}
        F_{I} &\sim \rho_l U_\mathrm{int}^2 A_\mathrm{int}, \label{eq:FI}\\
        F_{C} &\sim \sigma \pi d_B, \label{eq:FC}\\
        F_{B} &\sim \rho_l g \pi (A_\mathrm{int}^{1/2})^{3}/6. \label{eq:FB}
    \end{align}%
\end{subequations}

Among these three forces, $F_{I}$ acts as a driving force, while $F_{C}$ and $F_{B}$ act to resist the displacement of a bubble \cite{Talmor1977TwophaseMaps,Motil2003GasliquidMicrogravity}. Hence, we introduce two new dimensionless numbers:
\begin{subequations}\label{eq:Weber_numbers}\begin{align}
    We^* &= \frac{F_I}{F_C} = \rho_l U_\mathrm{int}^2 A_\mathrm{int}/\sigma \pi d_B, \label{eq:We*}\\
    \frac{We^*}{1+Bo^*} &= \frac{F_I}{F_C + F_B} = \frac{\rho_l U_\mathrm{int}^2 A_\mathrm{int}/\sigma \pi d_B}{1 + \rho_l g \pi A_\mathrm{int}^{3/2}/6\sigma \pi d_B}. \label{eq:We*ratio}
\end{align}\end{subequations}
The modified Weber number $We^*$ in Eq.~\eqref{eq:We*} captures the pore-scale force balance between $F_{I}$ and $F_{C}$. Meanwhile, the modified Bond number $Bo^*=\rho_l g \pi A_\mathrm{int}^{3/2}/6\sigma \pi d_B$ captures for the pore-scale force balance between $F_{B}$ and $F_{C}$. Then, the ratio between inertia and a combination of capillary and buoyancy forces at the pore scale can be evaluated by the dimensionless number $We^*/(1+Bo^*)$ introduced in Eq.~\eqref{eq:We*ratio}. Importantly, our choice of scales leads these ratios to naturally compare the forces that dynamically deform the bubble (inertia and buoyancy) to the force (capillarity) that seeks to hold the bubble static. Furthermore, we can define a capillary number in the usual way as $Ca = \mu_l U_{ls}/\sigma$, which is the ratio of the magnitude of the viscous force to the capillary force acting on a bubble. In our simulations, $Ca$ is on the order of $10^{-3}$; hence, viscous forces can be neglected in our analysis. 

We investigate the physics of bubble motion qualitatively by visualizing the bubble profiles from the volume fraction $\alpha_g(x,y,z,t)$ cut across the $(x,z)$ and $(y,z)$ planes. We supplement these cross-sections with 3D isometric views of the pore-scale resolved CFD simulations. We provide a quantitative understanding of the physics of bubble motion by tracking the evolution of $We^*$ and $We^*/(1+Bo^*)$ in time from the simulation data. Further, it must be noted that in the simulations presented below, the bubble is patched at the top center of the PBR and is displaced by the inlet water phase. However, we verified that the physics (and hence the results and discussion below) are independent of the initial bubble patching location.

\subsection{Bubble entrapment vs displacement: role of liquid inertia}

In this section, we discuss the influence of liquid inertia on the bubble dynamics under microgravity conditions by varying the global Weber number $We_{d_p}$ via the liquid phase's inlet superficial velocity $U_{ls}$. Since gravity is negligible in these conditions, it is expected that buoyancy force is also negligible, and hence its impact on bubble motion can be ignored. Consequently, the bubble displacement/entrapment is governed by the competition of the inertia and capillary forces. We perform simulations for three values of $We_{d_p}$ ($=0.165$, $0.42$, and $1.0$) to observe different pore-scale mechanisms set by the balance between the inertia and capillary forces. Since the magnitude, $F_{C}$, of the capillary force depends only on $\sigma$ and $d_B$ (recall Eq.~\eqref{eq:FC}), it can be considered constant for all the simulations. Meanwhile, since the magnitude, $F_{I}$, of the inertia force depends on $U_{ls}$ (recall Eq.~\eqref{eq:FI}), its magnitude varies in these simulations.

\begin{figure}
    \centering\includegraphics[width=\linewidth]{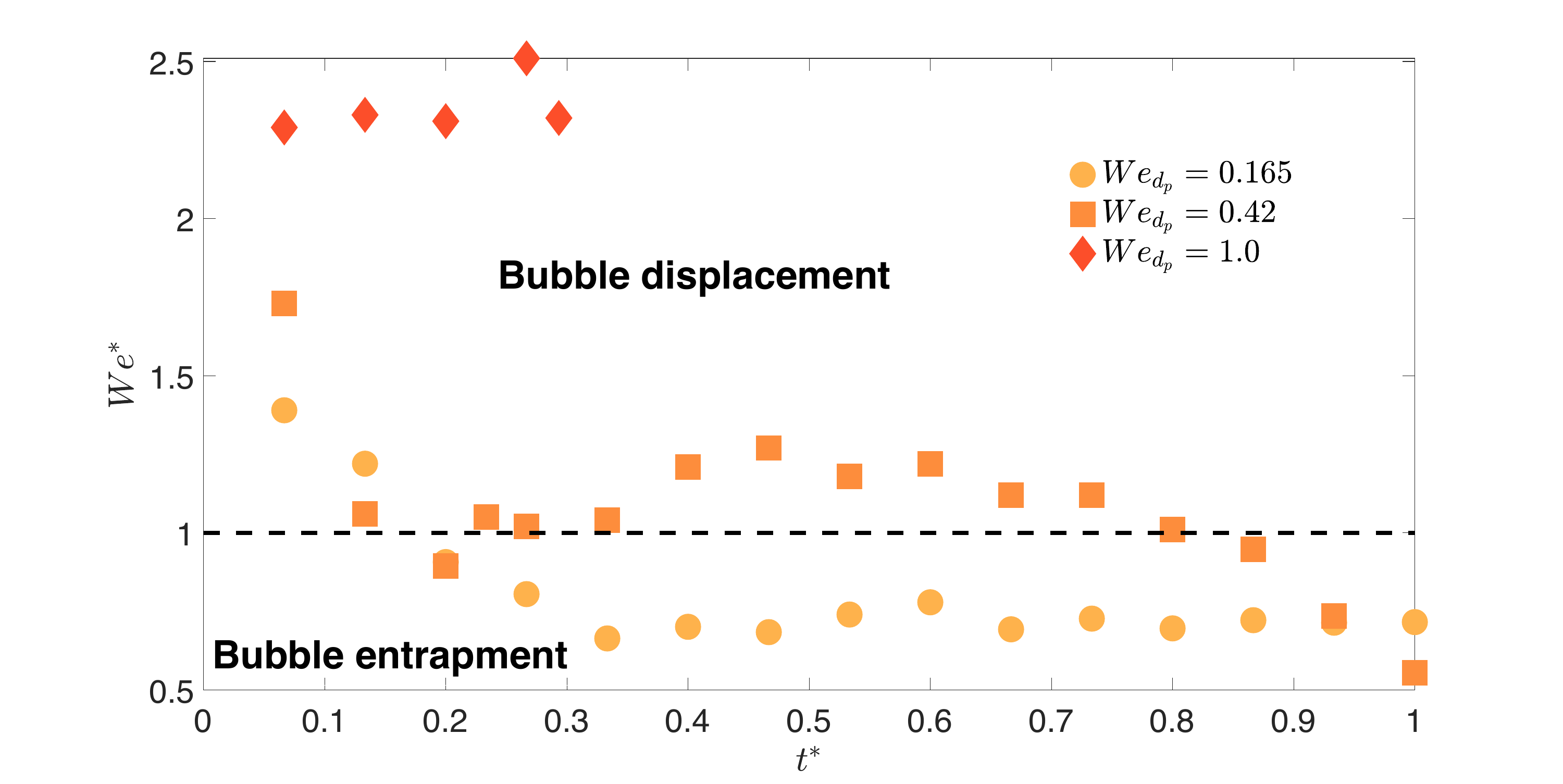}
    \caption{Evolution of $We^*=F_{I}/F_{C}$ demonstrates bubble displacement ($We^*>1$) at $We_{d_p} = 0.42$ and $We_{d_p} = 1.0$, while bubble entrapment ($We^*<1$) at $We_{d_p} = 0.165$.}
    \label{fig::FvsUls}
\end{figure}

For $We_{d_p} = 0.165$ in Fig.~\ref{fig::Uls_0.063ms_0g}, \emph{capillary entrapment} of the bubble is observed. As seen in the 2D planar cuts and 3D isometric view, at early times (small $t^*$), the bubble displaces slightly due to liquid inertia but later is entrapped indefinitely due to the dominance of $F_{C}$ over $F_{I}$. In fact, the bubble shape remains nearly spherical at all times shown due to the weak influence of $F_{I}$. This behavior is quantified in Fig.~\ref{fig::FvsUls}, wherein $We^*$ reduces below $1$ at early times (small $t^*$), which in turn explains why we observe capillary entrapment for $We_{d_p} = 0.165$ (a number that would otherwise not shed any light on whether or not to expect entrapment).

\begin{figure*}
    \centering
    \begin{subfigure}[t]{\linewidth}
      \includegraphics[width=\linewidth]{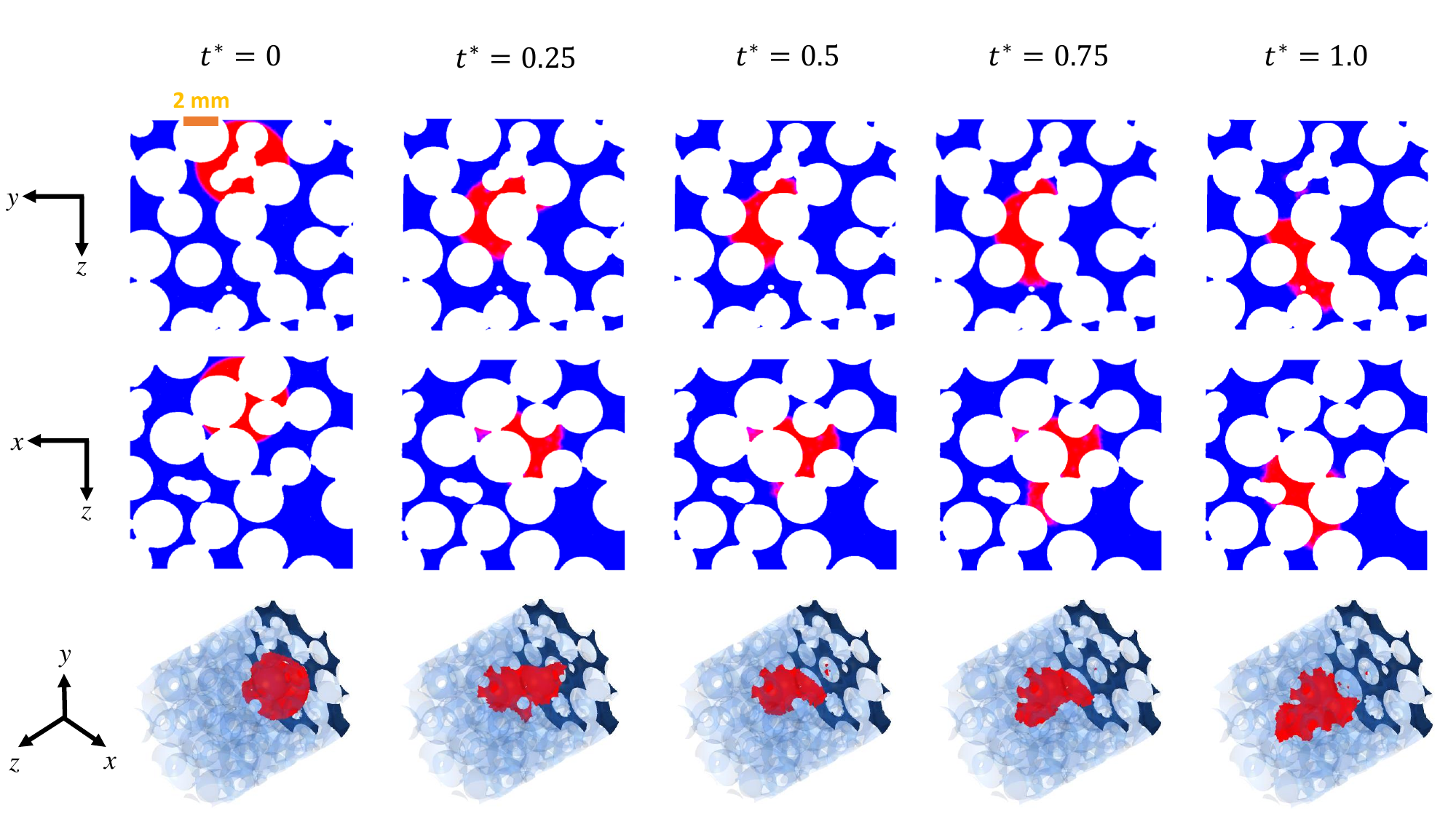}  
      \caption{}
      \label{fig::Uls_0.1ms_0g}
    \end{subfigure}
    \begin{subfigure}[t]{\linewidth}
      \includegraphics[width=\linewidth]{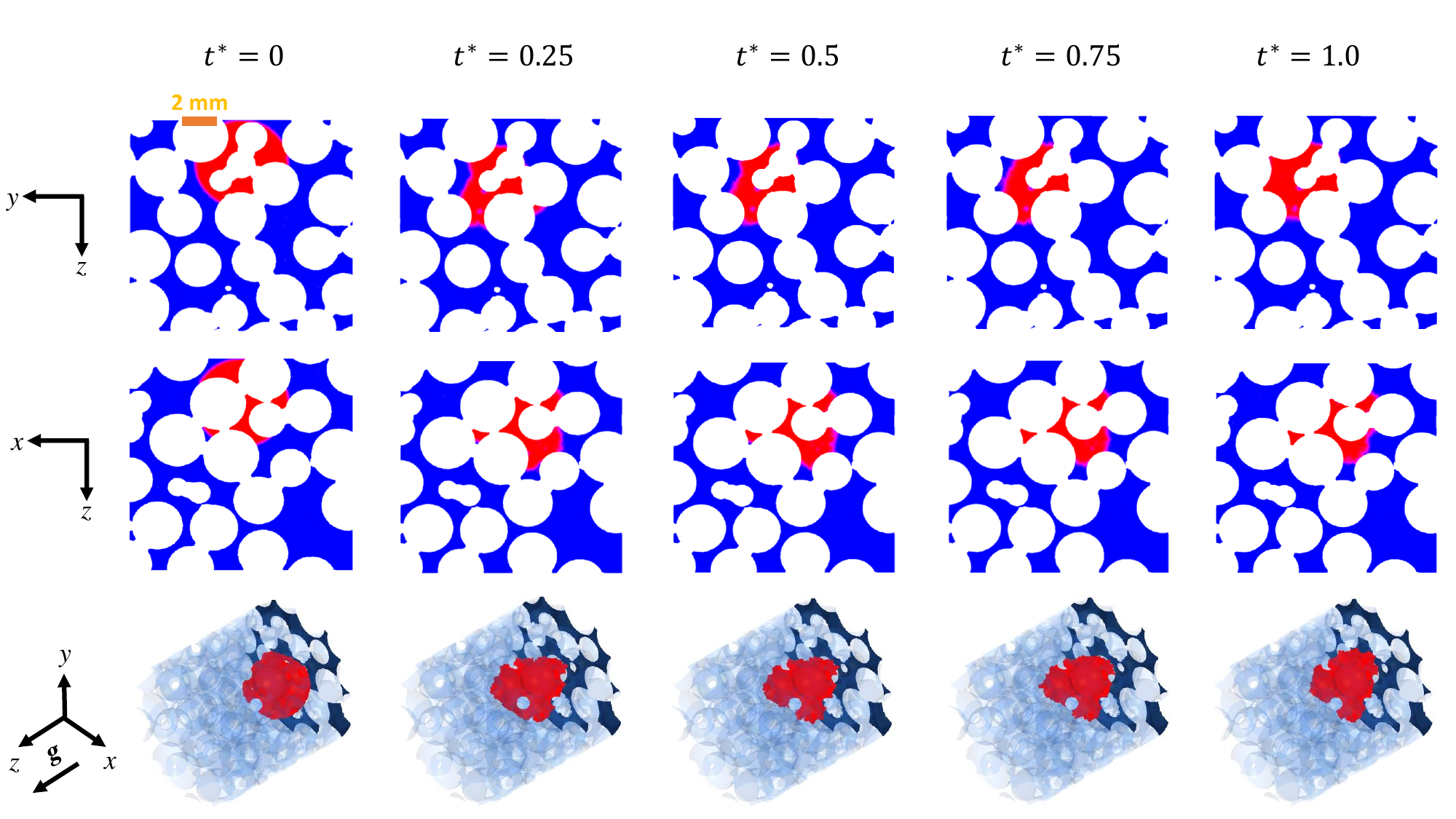}  
      \caption{}
      \label{fig::Uls_0.1ms_1g}
    \end{subfigure}
    \caption{Bubble evolution in 2D axial profiles (top and middle rows) and 3D isometric views (bottom rows) for $We_{d_p} = 0.42$ at (a) microgravity showing inertia-induced bubble displacement, followed by capillary entrapment and (b) earth's gravity showing buoyancy entrapment. The liquid flow and gravity are in the $+z$ direction.}
    \label{fig::Uls_0.1ms_0g_1g}
\end{figure*}

Next, for $We_{d_p} = 0.42$ in Fig.~\ref{fig::Uls_0.1ms_0g}, we observe bubble displacement due to the dominance of the inertia force, after which the bubble becomes stuck indefinitely as $t^* \to 1$ (as seen in the 2D planar cuts and 3D isometric view). The bubble undergoes deformation from its original spherical shape in order to traverse through the tortuous pore structures (visible in the isometric view). This behavior is further explained by Fig.~\ref{fig::FvsUls}, in which we observe that at early times (small $t^*$), $We^*>1$, indicating the dominance of the inertia force, which leads to bubble displacement. Then, at $t^*\approx0.85$, $We^*$ falls below $1$, indicating the dominance of the capillary force, which leads to bubble entrapment at later times. The pore-scale mechanism for these dynamics is that, as the bubble displaces through the PBR, $F_{I}$ decreases with time due to the combined reduction in gas interfacial area $A_\mathrm{int}$ and area-averaged interfacial velocity $U_\mathrm{int}$ (recall Eq.~\eqref{eq:FI}).

Finally, for $We_{d_p} = 1.0$ in Fig.~\ref{fig::Uls_0.155ms_0g}, the bubble readily displaces the water already in the PBR and is not entrapped. This phenomenon can be termed as \emph{inertia-induced bubble displacement}. Due to the dominance of $F_{I}$ in this case, the bubble elongates away from its spherical shape. Further, since $F_{I} \gg F_{C}$ in this case, the bubble eventually breaks up into smaller bubbles, as seen in the isometric view of Fig.~\ref{fig::Uls_0.155ms_0g}. The time taken for the bubble to displace is much shorter than in the cases above. In addition, we observe from Fig.~\ref{fig::FvsUls} that, at early times, $We^*$ is increasing. Indeed, a larger $We^*$ demonstrates the growing dominance of $F_{I}$ over $F_{C}$, which leads to the break up of the bubble. The pore-scale mechanism for this observation is that the interfacial area $A_\mathrm{int}$ is \emph{increasing} as the bubble distorts and elongates. The data for this case ends before $t^*=1$ because the broken bubble leaves the REV under consideration before $t^*=1$.

\subsection{Influence of gravity}
\label{sec:grav}

In this section, we evaluate the impact of gravity on the dynamics of a gas bubble traversing a PBR. The bubble flow regime is widely observed under both microgravity and terrestrial gravity conditions. We consider $We_{d_p} = 0.42$ with both $g=10^{-4}~\si{\meter\per\second\squared}$ and $g=9.8~\si{\meter\per\second\squared}$ to elucidate the influence of the buoyancy force on the bubble flow profiles. The buoyancy force, $F_{B}$, acts in the direction opposite of gravity. In our simulations, the flow direction is $+z$ and, hence, $F_{B}$ acts in the $-z$ direction. Thus, the buoyancy force opposes the bubble displacement so that $F_{B}$ aids $F_{C}$ to cause bubble entrapment.

Figure~\ref{fig::Uls_0.1ms_1g} illustrates the pore-scale resolved simulations of bubble flow at terrestrial gravity conditions. The bubble experiences \emph{buoyancy entrapment} because the resisting forces, $F_{C}$ and $F_{B}$, dominate the displacing force, $F_{I}$. Put quantitatively, $We^*/(1+Bo^*) < 1$. The resisting forces cause the bubble shape to be nearly spherical and arrest the displacement at the location where the bubble was initially patched at $t^*=0$. This behavior (Fig.~\ref{fig::Uls_0.1ms_1g}) is contrary to that in  Fig.~\ref{fig::Uls_0.1ms_0g} (for the same value of $We_{d_p} = 0.42$ but at microgravity conditions), because when buoyancy was negligible, the inertia force was initially stronger than the capillary force, leading to bubble displacement, though inertia was later overcome by capillarity leading to entrapment. 

\begin{figure}
    \centering\includegraphics[width=\linewidth]{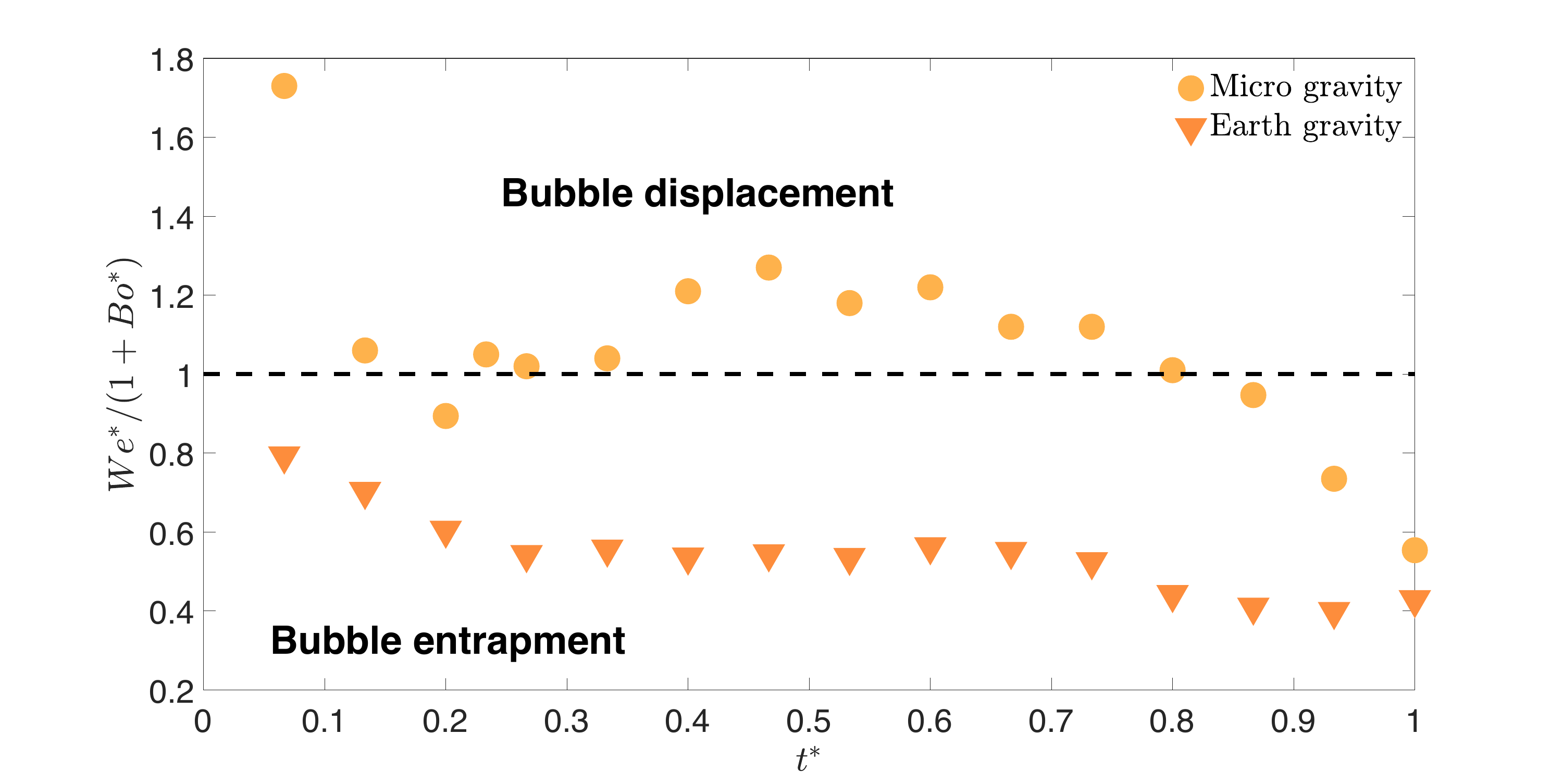}
    \caption{The evolution of $We^*/(1+Bo^*)$ for fixed $We_{d_p}$ and two values of $g$, shows the initial bubble displacement ($We^*/(1+Bo^*)>1$) at microgravity, but perpetual bubble entrapment ($We^*/(1+Bo^*)<1$) at earth gravity conditions.}
    \label{fig::Fvsg}
\end{figure}

Further, a quantitative comparison between the bubble dynamics under microgravity and earth gravity conditions for $We_{d_p} = 0.42$ is presented in Fig.~\ref{fig::Fvsg}. Due to the synergy between $F_{C}$ and $F_{B}$, the influence of $F_{I}$ is negligible, and hence the bubble is entrapped for $We_{d_p} = 0.42$ under terrestrial gravity conditions. This difference is quantitatively captured by the new dimensionless number $We^*/(1+Bo^*)$ introduced in Eq.~\eqref{eq:We*ratio}, which is always $<1$ for earth gravity conditions in Fig.~\ref{fig::Fvsg}, while it starts off $>1$ (displacement) before decreasing to a value $<1$ (entrapment) under microgravity conditions. 

\subsection{Bubble-to-pulse flow regime transition}
Under terrestrial gravity conditions, the primary flow transition is from a trickle to a pulse regime. Since the trickle flow regime has not been observed under microgravity conditions, the main flow regime transition observed in microgravity is from bubble to pulse flow. In experiments, Motil, Balakotaiah, and Kamotani \cite{Motil2003GasliquidMicrogravity} observed that this flow regime transition occurs when the gas flow rate is increased for a fixed liquid flow rate (assuming it is above a certain threshold). This observation can be explained by noting that increasing the amount of gas in the PBR causes the coalescence of existing bubbles and extends their reach across the width of the PBR. Prior experimental studies \cite{Motil2003GasliquidMicrogravity,Motil2021GasliquidExperiment,Taghavi2022GasliquidExperiment2} inferred the bubble-to-pulse transition by noting sudden jumps in the pressure oscillations measured by a downstream transducer, which they took as a sign indicating the formation of a pulse. In this section, we leverage our CFD simulations to uncover the pore-scale dynamics, at least qualitatively using volume fraction maps, of the two bubbles as they coalesce along the radial direction to form a pulse spanning the width of the bed.

\begin{figure*}
    \centering\includegraphics[width=\linewidth]{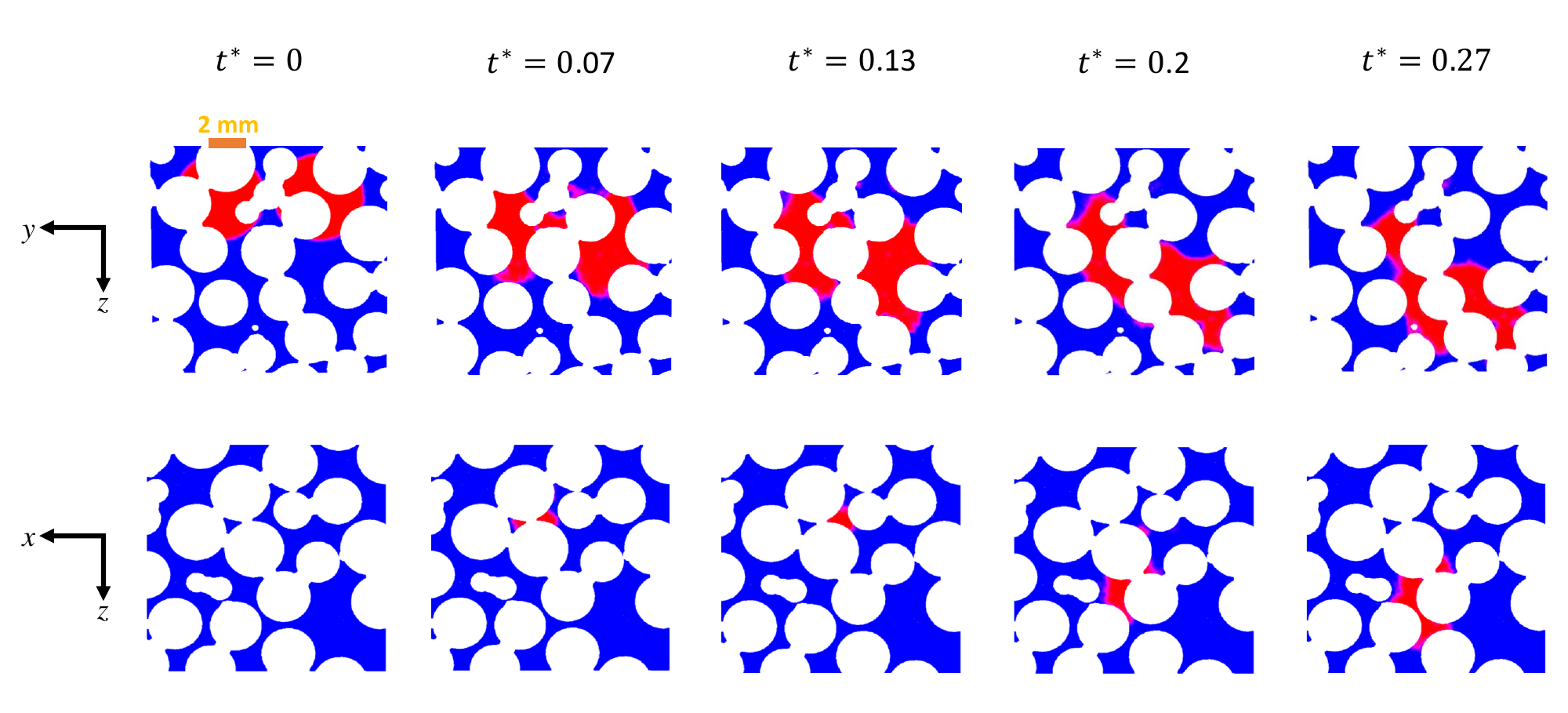}
    \caption{Bubble to pulse flow regime transition for $We_{d_p} = 0.6$ at microgravity displayed in 2D axial profiles. The liquid flow and the force of gravity are in the $+z$ direction.}
    \label{fig::bubble-to-pulse}
\end{figure*}

The formation of a pulse usually begins when two bubbles navigating through the pore spaces, predominantly along the axial (flow-wise) direction, meet and coalesce in the cross-section. To this end, we set up a simulation with $We_{d_p}=0.6$ in which we patched two bubbles of $5~\si{mm}$ diameter (see Fig.~\ref{fig::bubble-to-pulse}) near the entrance of the PBR at $t^*=0$. Since the flow is simulated within an REV, we only considered the case of two bubbles. In an actual PBR, there could be multiple bubbles merging to form a pulse spanning the width of the bed.
Next, as seen in Fig.~\ref{fig::bubble-to-pulse}, at $t^*\approx0.07$ the bubbles deform and flow through the pores and start interacting at $t^*\approx0.13$. The two bubbles coalesce to form a pulse at $t^*\approx0.2$. Subsequently, as seen in the plots at $t^*=0.27$, the pulse propagates downstream through the initially liquid-filled regions. Thus, we observe alternate gas-rich and liquid-rich regions. In an actual reactor, multiple such pulses are expected to form (for a sufficiently high gas flow rate); hence, recurring pulses (alternating gas-rich and liquid-rich regions) would be observed.

\section{Conclusion}
\label{sec::Conclusion}

Experiments on gas-liquid flow through a packed-bed reactor performed under microgravity conditions at the International Space Station showed the presence of the so-called bubble and pulse flow regimes but noted the absence of the so-called trickle and spray flow regimes \cite{Motil2003GasliquidMicrogravity,Motil2021GasliquidExperiment,Taghavi2022GasliquidExperiment2}. They found that the flow regime maps in microgravity differ from those under terrestrial gravity conditions, leaving a knowledge gap regarding the understanding of how gravity shapes the presence (or absence) of regimes in gas-liquid flow through a porous medium. High-fidelity simulations can bridge this gap. To this end, in this work, we performed interface-resolved CFD simulations to study the microgravity bubble flow regime in a packed-bed reactor, both qualitatively and quantitatively.

First, we described and implemented a workflow for generating packed-bed geometries using rigid-body simulations performed in the rendering software Blender. Second, from these geometries, we extracted a representative-volume element of a packed bed to use for CFD simulations. We used a shrink-wrapping algorithm to generate high-quality meshes of the flow domains. Third, we ran simulations to determine the impact of the packing-particle-diameter-based (global) Weber number, $We_{d_p}$, and gravity conditions on bubble entrapment and displacement. 

In our quantitative simulations at  $We_{d_p} = 0.165$ and $We_{d_p} = 0.42$, we observed capillary entrapment at early and late times, respectively, due to the dominance of the capillary force (which resists the bubble displacement) over the inertia force (which aids bubble displacement). Further, at $We_{d_p} = 0.165$, the bubble remains nearly spherical at all times. Meanwhile, at $We_{d_p} = 0.42$, the bubble undergoes deformation at early times, after which the bubble is entrapped. On the other hand, in the simulation for $We_{d_p} = 1.0$, the inertia force dominates the capillary force, resulting in inertia-induced bubble displacement. In this case, the bubble undergoes elongation and eventually breaks up. Moreover, for $We_{d_p} = 0.42$ under terrestrial conditions, since gravity acts in the flow direction, the buoyancy force causes additional resistance, which leads to buoyancy entrapment (as opposed to the bubble displacement followed by capillary entrapment that we observed under microgravity conditions). 

To unify our understanding of the pore-scale mechanisms that set the flow regime(s) observed in the simulations, we introduced pore-scale (local) dynamic scales, dependent on bubble interfacial area $A_\mathrm{int}$. Using the proposed dynamic scales, we estimated the magnitudes of the inertia and buoyancy forces, and we were able to explain several of the unsteady features observed via force balance arguments. Specifically, we introduced the new dimensionless numbers given in Eq.~\eqref{eq:Weber_numbers}, which depend on time. These modified Weber-like numbers allowed us to rationalize the dynamics of displacement and entrapment of bubbles in the bubble flow regime. Specifically, for the simulations at microgravity conditions, we obtained $We^* < 1$, $We^*\simeq1$, and $We^*\gg1$ for $We_{d_p}=0.165$, $We_{d_p}=0.42$, and $We_{d_p}=1.0$, respectively, demonstrating that $We^*$ delineates bubble entrapment from bubble displacement. Likewise, for the simulation at earth's gravity conditions, $We^*/(1+Bo^*)$ delineates bubble entrapment from bubble displacement. Specifically, our simulations indicated bubble entrapment at $We_{d_p}=0.42$ for which $We^*/(1+Bo^*)<1$.

Using a representative-volume element reduced the computational efforts required to simulate the bubble dynamics and made our interface-resolved study of the gas-liquid flow through a PBR possible. Nevertheless, the wall-clock computational times of our CFD simulations were on the order of two weeks. Thus, there is still room for improvement and a demonstrated need for reduced-order models (either data-driven or otherwise), such as 1D two-fluid models \cite{Salgi2014ImpactBeds,Salgi2015PulseConditions,Salgi2017Experimentally-basedBeds,Taghavi2019GasRate,Zhang2017HydrodynamicsModel}, to quantitatively investigate the pulse flow regimes and pressure drop correlations. This direction is left for future work. 

%%%%%%%%%%%% BACK MATTER %%%%%%%%%%%%

\clearpage

\section*{Author Contributions}
\textbf{Pranay P.\ Nagrani}: conceptualization (equal); data curation (lead); formal analysis (lead); investigation (equal); methodology (equal); validation (lead); visualization (lead); writing - original draft (lead).
\textbf{Amy M.\ Marconnet}: writing - review \& editing (supporting); supervision (equal); funding acquisition (supporting).
\textbf{Ivan C.\ Christov}: conceptualization (equal); formal analysis (supporting); investigation (equal); methodology (equal); writing - original draft (supporting); writing - review \& editing (lead); supervision (equal); funding acquisition (lead).

\section*{Acknowledgements}
This research was supported by the National Aeronautics and Space Administration under Grant No.\ 80NSSC22K0290.
%issued through the Glenn Research Center. 
Simulations were performed using the community clusters of the Rosen Center for Advanced Computing at Purdue University.

\section*{AUTHOR DECLARATIONS}
\subsection*{Conflict of Interest}
The authors have no conflicts to disclose.

\section*{Data Availability and Reproducibility Statement}
The data supporting this study's findings are openly available in the Purdue University Research Repository (PURR) at \url{http://doi.org/10.4231/2H02-XA61}. 

The simulation data, from which Figs.~\ref{fig::Uls_0.1ms_0.155ms_0g}, \ref{fig::Uls_0.1ms_0g_1g}, and \ref{fig::bubble-to-pulse} were generated, are provided in the PURR as five zip archives named for each global Weber number, $We_{d_p}$, described in the text. Each zip archive contains ANSYS Fluent 2022R1 case files and selected output data files.  Figures~\ref{fig::PBR_Geom_PBR}, \ref{fig::PBR_Geom_Mesh}, and \ref{fig::BC} can also be generated from these ANSYS files. The simulation data were validated via a grid-convergence study as described in the main text.

Another zip archive provided in the PURR includes a spreadsheet with the numerical data from Figs.~\ref{fig::PBR_Geom_Val}, \ref{fig::FvsUls}, and \ref{fig::Fvsg}, MATLAB scripts for generating these figures, as well as Blender 2.92.0 simulation files, and ANSYS SpaceClaim geometry files (.scdoc and .fmd). Blender was used to generate the packed-bed geometry, which was cleaned up and made ready using SpaceClaim for the Fluent meshing and simulations. The mesh used is also provided in the archive as an .msh file.

Finally, the Excel spreadsheet with the numerical data from Figs.~\ref{fig::PBR_Geom_Val}, \ref{fig::FvsUls}, and \ref{fig::Fvsg} is also made available as Supplementary Material for this article. This data was generated by post-processing the ANSYS Fluent simulations.

\section*{ORCiD}
\noindent
{\footnotesize\textit{Pranay P.\ Nagrani}~\orcidlink{0000-0003-4568-9318}~\url{https://orcid.org/0000-0003-4568-9318}}\\
{\footnotesize\textit{Amy M.\ Marconnet}~\orcidlink{0000-0001-7506-2888}~\url{https://orcid.org/0000-0001-7506-2888}}\\
{\footnotesize\textit{Ivan C.\ Christov}~\orcidlink{0000-0001-8531-0531}~\url{https://orcid.org/0000-0001-8531-0531}}

%\section*{References}
%%  whoever last updates the linked Mendeley .bib changes the line below,
%% "pbr.bib" is Pranay and "pbr2.bib" is ICC
\bibliography{pbr2.bib}

\end{document}